\documentclass[12pt,twoside]{article}
\usepackage{fleqn,espcrc1}

\usepackage{graphicx}

\newcommand{\AmS}{{\protect\the\textfont2
  A\kern-.1667em\lower.5ex\hbox{M}\kern-.125emS}}

\title{Strangeness production in heavy ion collisions}

\author{Krzysztof Redlich\address{
        Gesellschaft f\"ur Schwerionenforschung, D-64291 Darmstadt, Germany
        }%
        \thanks{On leave of absence from: Institute  of
        Theoretical Physics, University of Wroc\l aw, PL-50204
        Wroc\l aw, Poland.
                }\thanks{Work supported in part by: Committee for Scientific Research (KBN-2P03B 03018) }}

\begin{document}
\maketitle
\begin{abstract}
Strangeness   production in heavy ion collisions is discussed in
a broad energy range from SIS to RHIC. 
In  the whole energy  range  particle yields are showing  high
level of chemical equilibration  which can be described by the
unified freezeout conditions of fixed energy/particle $\simeq
1$GeV. The statistical model within the canonical formulation of
strangeness conservation provides  a framework to describe  the
observed enhancement of (multi)strange particles  from p+A to A+A
collisions measured at the SPS energy and predicts that this
enhancement should be larger for decreasing collision energy.
However, only at the SPS and RHIC chemical freezeout temperature
is consistent within error with the critical value required for
deconfinement and simultaneously strangeness is uncorrelated and
distributed in the whole volume of the fireball.
\end{abstract}

\section{Introduction}

Ultrarelativistic heavy ion collisions provide a unique
opportunity to study the properties of  highly excited hadronic
matter under extreme conditions of high density and high
temperature \cite{qm,satz,stachel,stoc,heinz1}. From the analysis
of rapidity distribution of protons and of their transverse energy
measured in 158 A GeV/c Pb+Pb collisions an estimate of the
initial conditions  \cite{satz,stachel,stoc,heinz1} leads to an
energy density of 2-3 GeV/fm$^3$ and a baryon density  of the
order of 0.7/fm$^3$. Lattice QCD at vanishing baryon density
suggests that the phase transition from confined to the
quark-gluon plasma (QGP) phase appears at the temperature
$T_c=173\pm 8$ MeV which corresponds to the critical energy
density $\epsilon_c\sim 0.6\pm 0.3$ GeV/fm$^3$ \cite{karsch}. One
could thus conclude that the required initial conditions for quark
deconfinement are already reached in heavy ion collisions at the
SPS energy. It will be discussed to what extent the composition of
the final-state hadrons, in particular their strangeness content
can be considered as a probe of quark deconfinement in the initial
state.
\begin{figure}[htb]
\vskip -0.3cm
\begin{minipage}[t]{75mm}
\includegraphics[width=17.9pc, height=12.4pc]{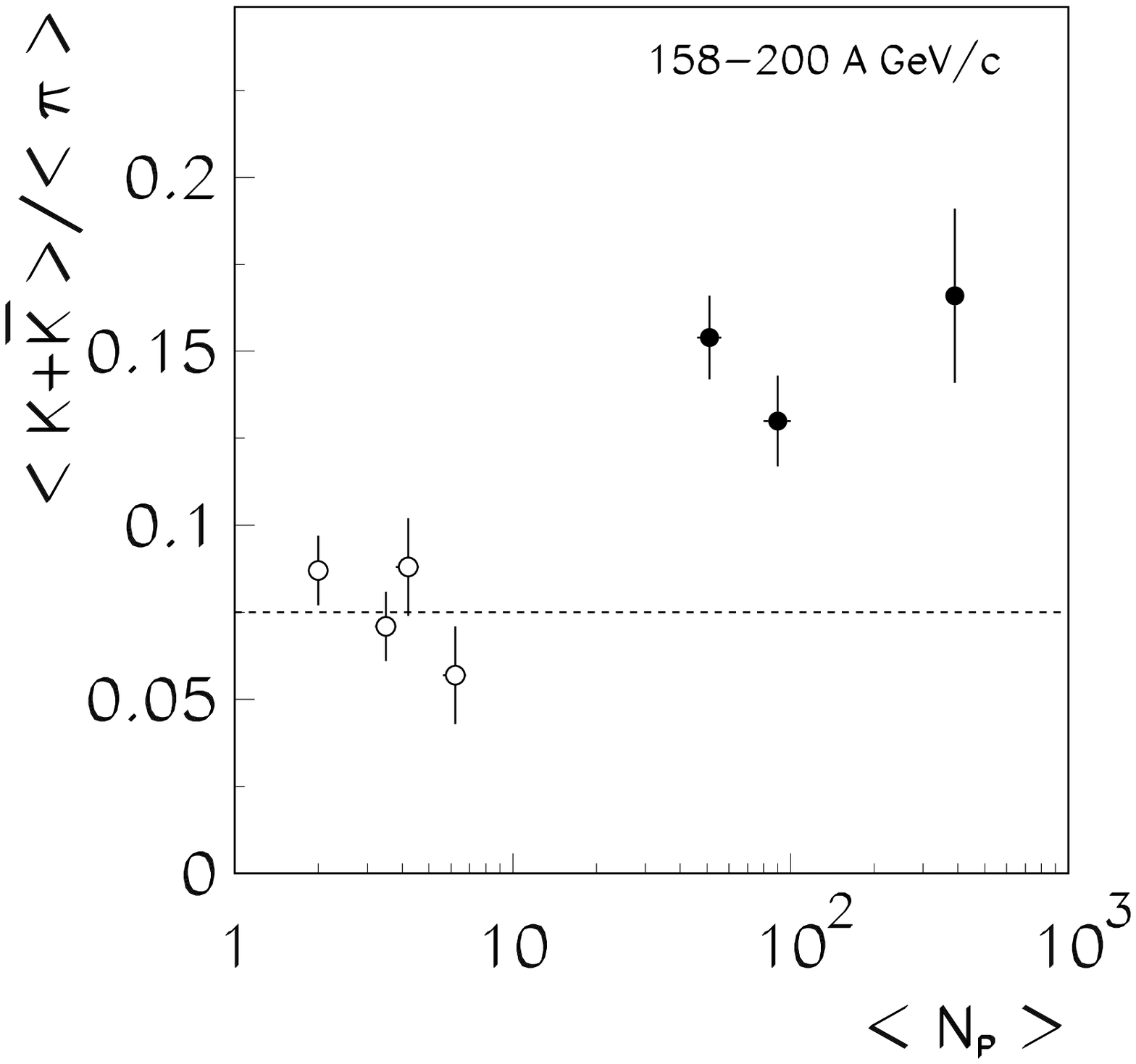}\\
\vskip -1.0true cm \caption{ Ratio of total number of kaons per
pions versus the number of participant in p+p, p+A and A+A
collisions \cite{bialkowska}.}
\label{fig:1}
\end{minipage}
\hspace{\fill}
\begin{minipage}[t]{75mm}
\includegraphics[width=17.5pc,  height=12.4pc]{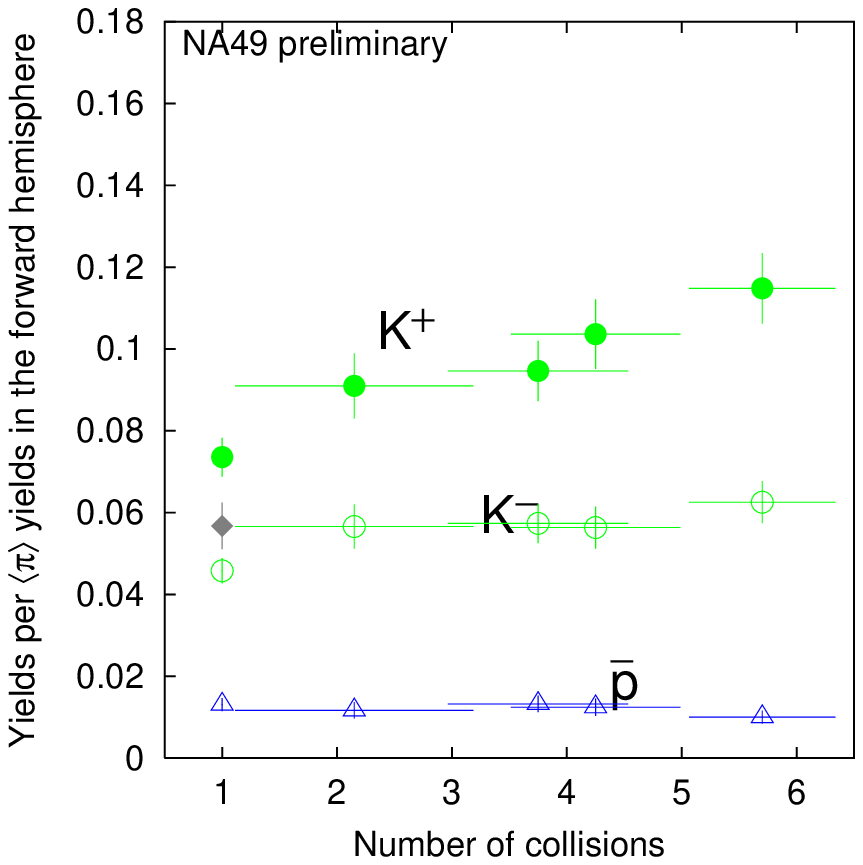}\\
\vskip -1.0 true cm \caption{The multiplicity  of $K^+,K^-$ and
$\bar p$ per pion multiplicity  as a function of the number of
collisions in p+A reactions \cite{sikler}.} \label{fig:2}
\end{minipage}
\end{figure}
\section{Strangeness content of the quark-gluon plasma - signal for deconfinement}
An enhanced production of strange particles was long suggested as
a possible signal of the QGP  formation in heavy ion collisions
\cite{rafelski1,letessier}. In the QGP the production and
equilibration of strangeness is very efficient due to  a large
gluon density and a low energy threshold for dominant QCD
processes of  $s\bar s$ production. In  hadronic systems the
higher threshold for strangeness production  was argued
\cite{rafelski1} to make the strangeness yield considerably
smaller and the equilibration time much longer.\footnote{ In
\cite{greiner} it was argued, however, that multi-mesonic
reactions could accelerate the equilibration time of strange
antibaryons especially when the  hadronic system is hot and very
dense.}

On the basis of the above strangeness QGP  characteristics  the
following experimental implications are commonly quoted as the
signal of deconfinement:

i) {\it global strangeness enhancement}: strangeness content of
secondaries,  measured  by the total number of produced $<s\bar
s>$ quarks per participant $A_{part}$ or per produced light quarks
$<u\bar u +d\bar d>$ should increase from pp, pA to AA collisions.

ii) {\it enhancement of multistrange baryons}: the specific
enhancement (increasing with strangeness content)  of
multi-strange baryons and anti-baryons in central AA collisions,
with respect to proton induced reactions  follows deconfinement.

iii) {\it chemical equilibration of secondaries}: the appearance
of the QGP being close to chemical equilibrium and subsequent
phase transition should   in general  drive hadronic constituents
produced from hadronizing QGP  towards chemical equilibrium.
\begin{figure}[htb]
 {\hskip -.1cm
\includegraphics[width=35.5pc, height=11.9pc]{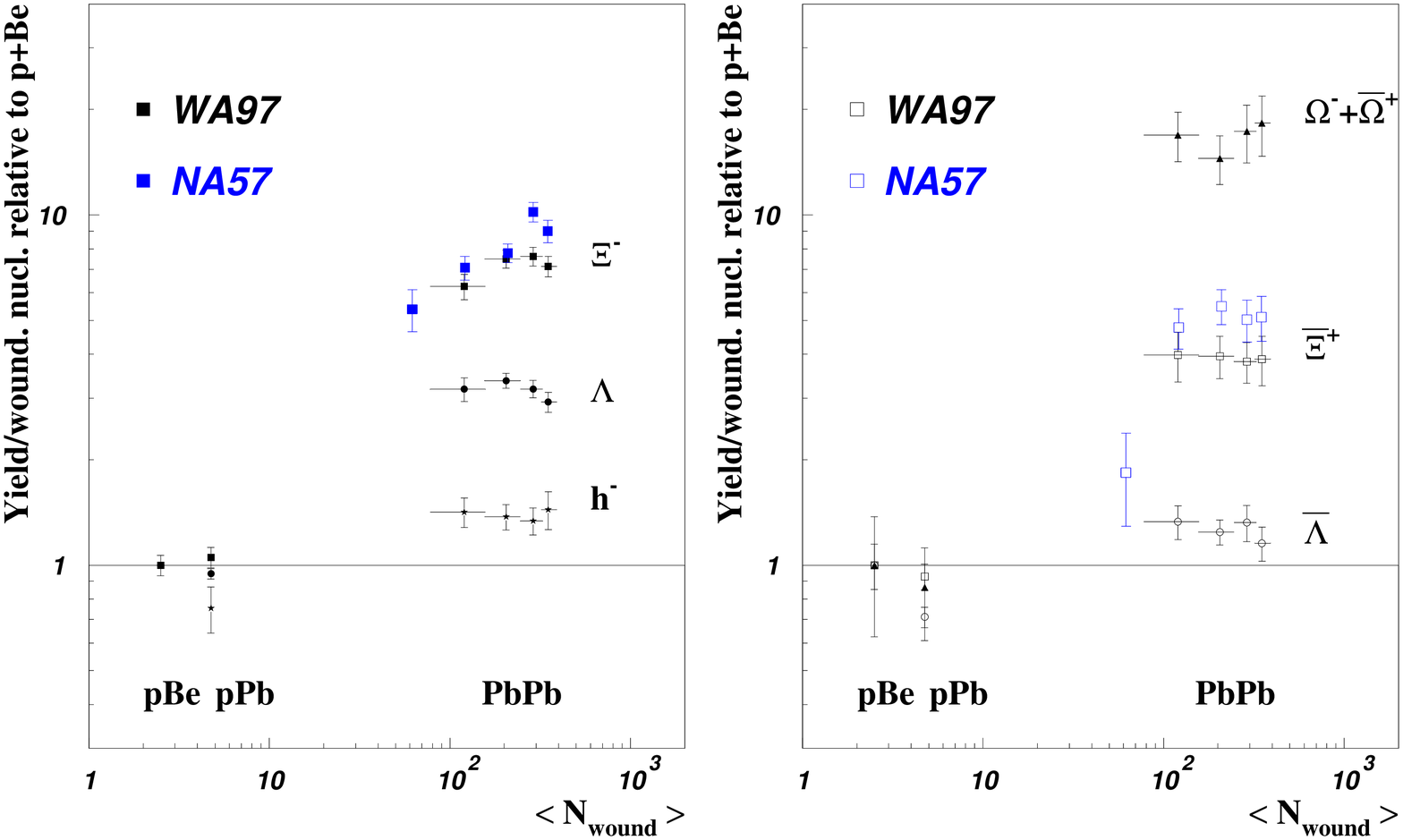}}\\
\vskip -1.0 true cm \caption{Particle yields per participant in
Pb+Pb relative to p+Be and p+Pb collisions centrality dependence.
The data are from WA97 \cite{andersen} and NA57 \cite{carrer}
collaboration.} \label{fig:3}
\end{figure}

Heavy ion experiments at CERN SPS  reported actually a global
strangeness enhancement   from p-p, p-A to A-A collisions
\cite{qm,sikler,bialkowska,andersen}. In Fig. 1 we show full
phase-space data of NA35 and NA49 collaboration on $<K+\bar
K>/<\pi>$ ratio in S+S, S+Ag and Pb+Pb  relative to p+p and p+A
collisions. There is indeed  a jump of factor of about two in
strangeness content when going from p+p to heavy ion collisions.
However, strangeness enhancement is already present in p+p
\cite{satz} and in p+A collisions \cite{sikler}. In Fig. 2 the
$<K^+>/<\pi^+>$ ratio measured by the NA49 collaboration in p+Pb
collisions  in the forward hemisphere shows an increase with the
number of collisions due to the secondary production. This effect
is qualitatively similar to the one observed  in Pb+Pb collisions.
Strangeness enhancement from p+p to the most central p+A
collisions can already account for more than 50$\%$ increase of
strangeness seen in Fig. 1.

Large strangeness content of the QGP plasma should be reflected in
a very specific hierarchy of multistrange baryons
\cite{rafelski1}: enhancement of $\Omega > \Xi >\Lambda$. Fig. 3
shows the yield/participant in Pb+Pb relative to p+Pb collisions
measured by the WA97 and NA57  collaboration
\cite{andersen,carrer}. Indeed the enhancement pattern of the
(anti)hyperon yields is seen to increase with strangeness content
of the particle and there is a saturation of this enhancement for
$N_{wond} > 100$.  Recent results  of the NA57 collaboration are
showing in addition an abrupt  change of anti-cascade enhancement
for a lower centrality. Similar behavior  was previously seen on
the level of $K^+$ yield measured by the NA52 experiment in Pb-Pb
collisions \cite{kabana}. These results are very interesting as
they might be interpreted as an indication of the onset of a new
dynamics. However, a more detailed experimental study and
theoretical understanding are still required here. It is e.g. not
clear  what is an origin of different centrality dependence of
$\Xi$ and $\bar\Xi$ as well as inconsistences   between the NA52
and the WA97 data.
\begin{figure}[h]
\vskip -1.5 cm
\begin{minipage}[t]{80mm}
\includegraphics[ width=19.5pc, height=14.7pc]{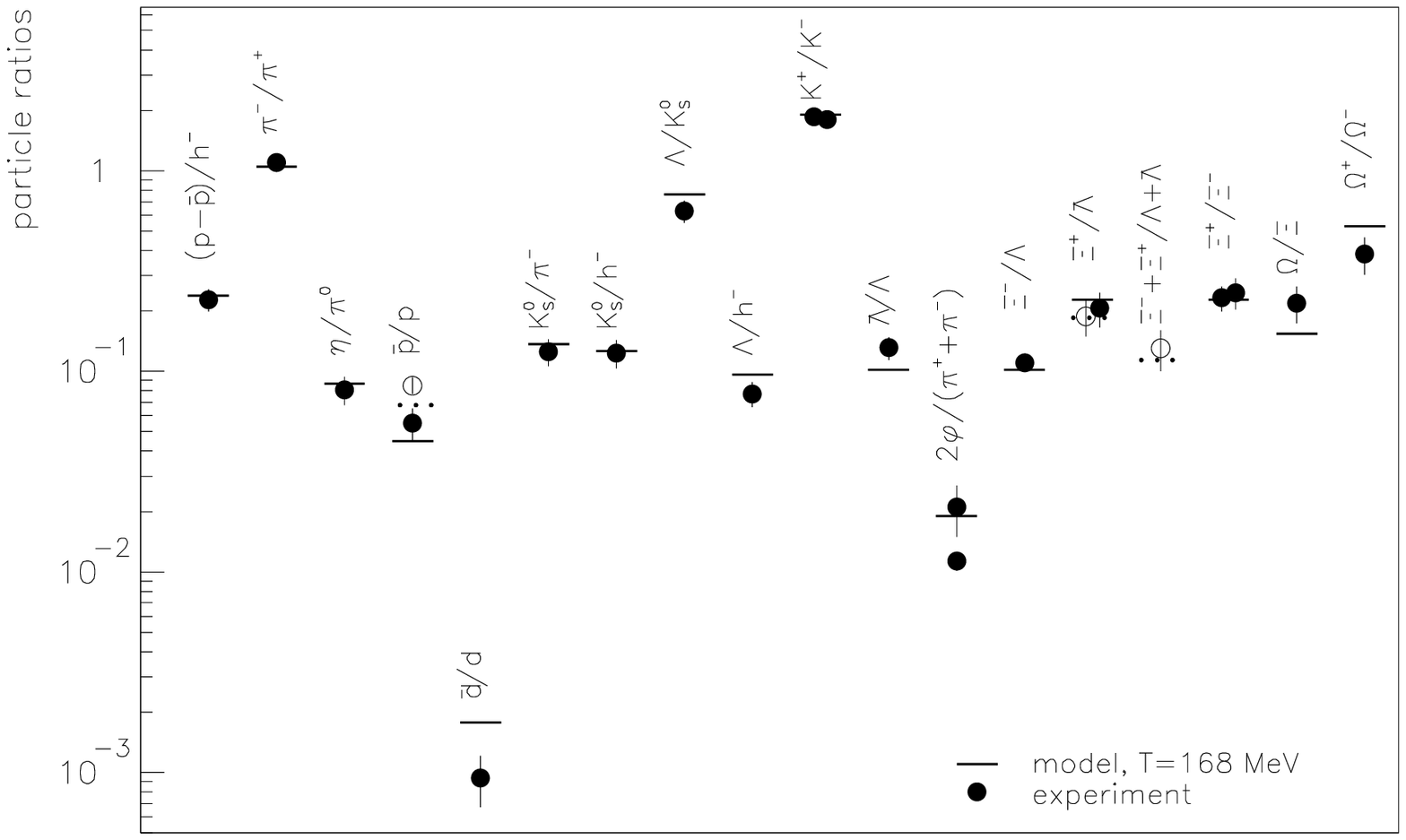}\\
\vskip -2. true cm \caption{The experimental data for  particle
multiplicity ratios  at the SPS (full points). The predictions of
thermal model (horizontal lines) \cite{braun}. } \label{fig:4}
\end{minipage}
\hspace{\fill}
\begin{minipage}[t]{75mm}
\includegraphics[width=17.9pc,height=15.3pc]{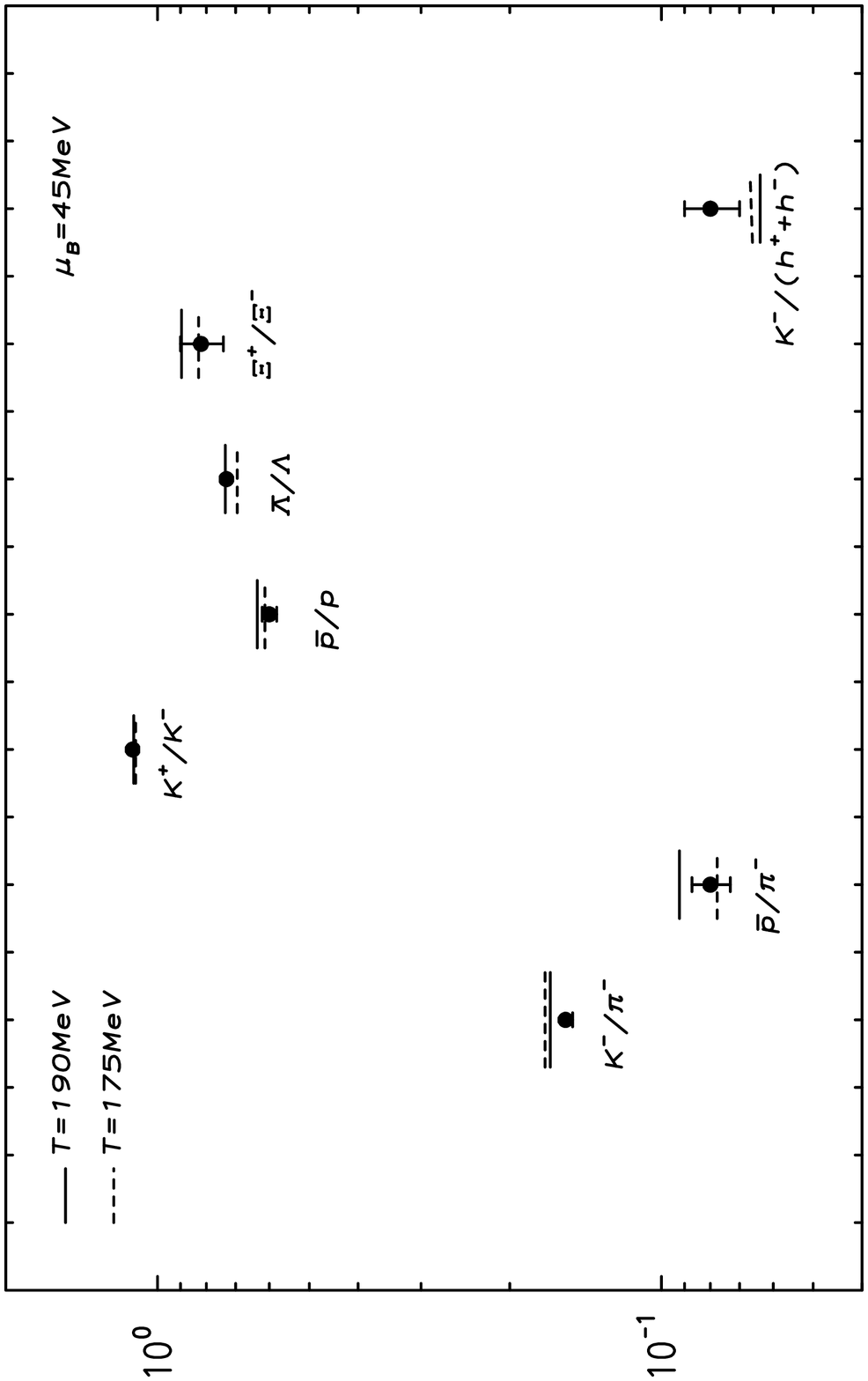}\\
\vskip -2. true cm \caption{Particle multiplicity ratios from STAR
collaboration at RHIC (full points) \cite{harris}. The thermal
model results are shown by the horizontal lines.} \label{fig:5}
\end{minipage}
\end{figure}

A number of different mechanisms was considered   to describe the
magnitude of the enhancement and centrality dependence of
(multi)strange baryons measured by the WA97
\cite{soff,bravina,vance,bleicher,capella,lin}. Microscopic
transport models make it clear that the data shown in Fig. 3 can
not be explained by pure final state  hadronic interactions. Only
the combination of the formers with an additional pre-hadronic
mechanisms    like baryon junction processes, color ropes or color
flux tubes overlap can partly explain the enhancement pattern and
the magnitude for the most central collisions. However, the
detailed centrality dependence is still not well reproduced within
the microscopic models. An alternative description of multistrange
particle production was formulated in terms of macroscopic thermal
models \cite{letessier,braun,hamieh} and will be discussed in the
next sections.

\section{Equilibration and particle yields}

One of the key questions in the description of particle production
in heavy ion collisions is to what extent measured  yields are
showing  equilibration. The level of equilibration of secondaries
in heavy ion collisions can be tested on  a different level:  by
analyzing (i) particle abundances or (ii) particle momentum
distributions. In the first case one establishes the chemical
composition of the system ({\it chemical freezeout}), while in the
second an additional information on dynamical evolution and
collective flow can be extracted ({\it thermal freezeout}).
%
\begin{figure}[htb]
\vskip -1.cm
\begin{minipage}[t]{80mm}
\includegraphics[width=17.5pc, height=16.5pc]{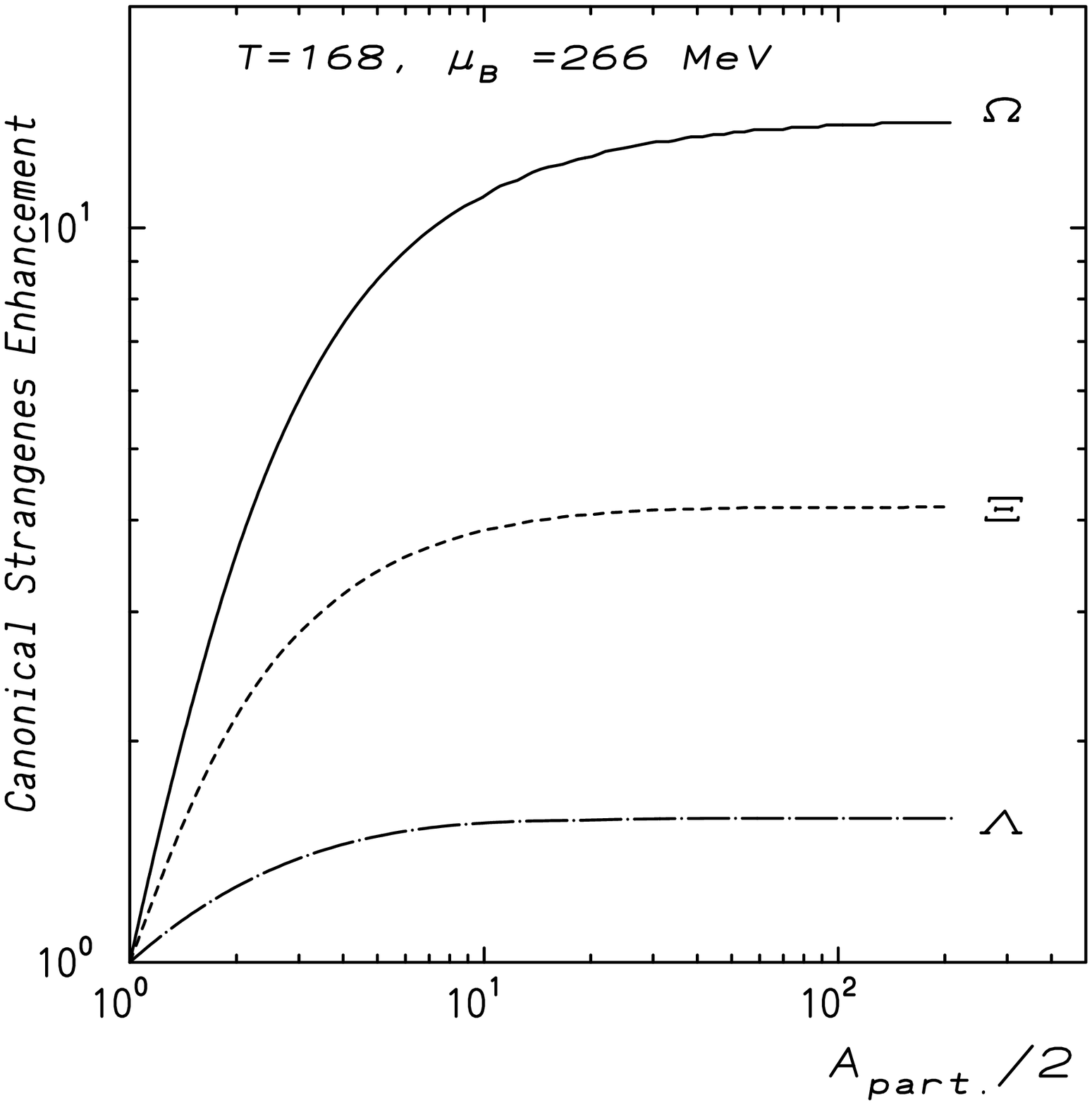}\\
\vskip -2. true cm \caption{Particle multiplicities per
participant normalized to its value in p+p system as a function of
$A_{part.}$  calculated  in statistical model in (C) ensemble
\cite{hamieh}. } \label{fig:6}
\end{minipage}
\hspace{\fill}
\begin{minipage}[t]{75mm}
\includegraphics[width=17.5pc, height=16.5pc]{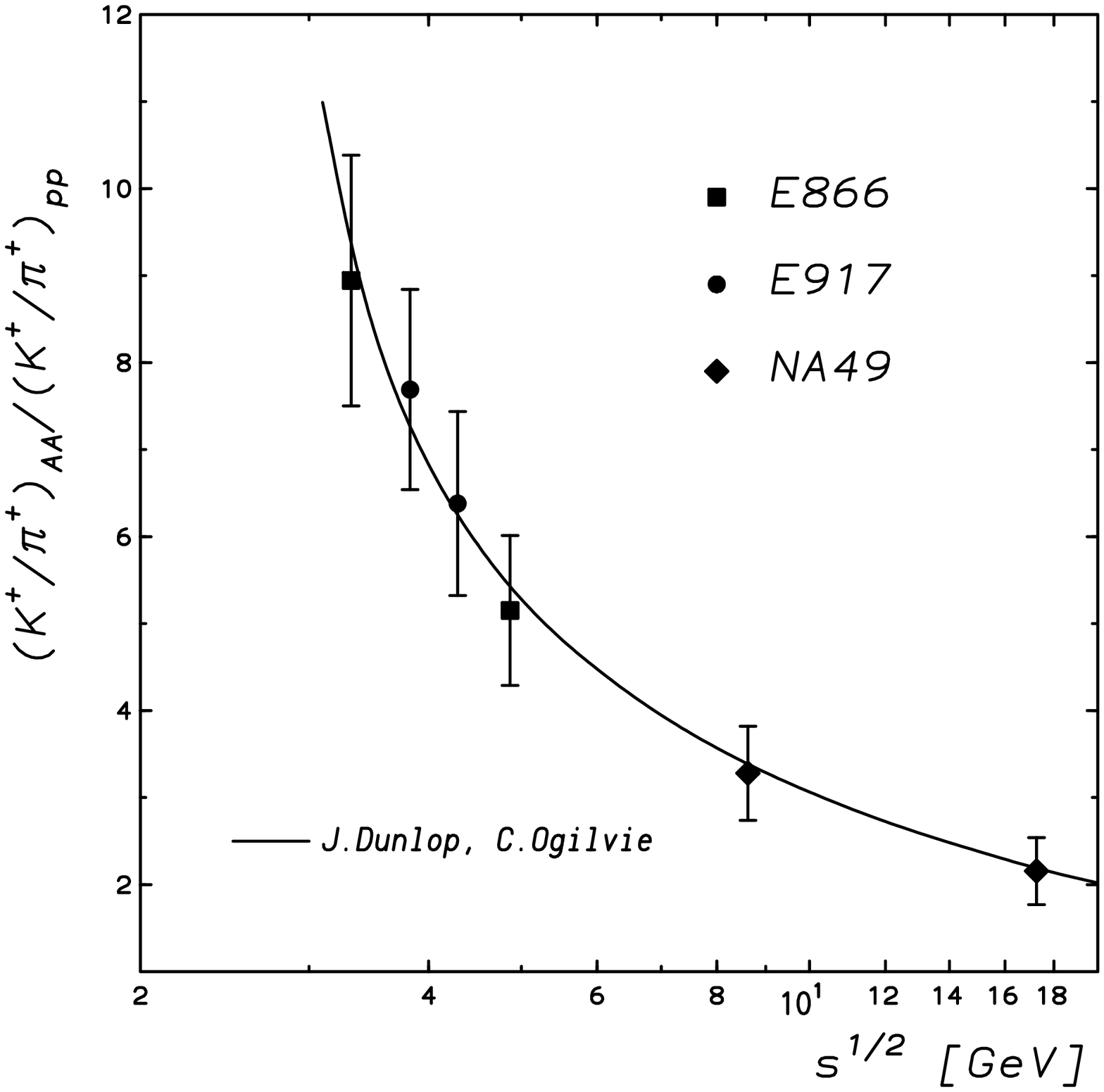}\\
\vskip -2. true cm \caption{ $K^+/\pi^+$ ratio at midrapidity in
A+A relative to p+p collisions. For the  compilation of data see
\cite{ogilvie,blume}. The full line is the parameterization from
\cite{ogilvie}.}
 \label{fig:7}
\end{minipage}
\end{figure}

The compilation of the experimental data for particle multiplicity
ratios in central Pb+Pb collisions at the SPS is shown in Fig. 4.
The relative hadron abundances are compared with the thermal model
\cite{braun}. The model was formulated  in the grand canonical
(GC) ensemble with the partition function  which contains  the
contributions of   most hadrons and resonances and preserves the
baryon number, strangeness and charge conservation. The particle
ratios depend only on two  parameters;  temperature $T$ and baryon
chemical potential $\mu_B$. It is clear from Fig. 4 that with
$T\sim 170$MeV, corresponding to the  energy density
$\epsilon_f\sim 0.6$GeV/$fm^3$, and with  the baryon chemical
potential $\mu_B\sim 270$MeV the statistical model  reproduces the
experimental data. 

In Fig. 5 we compare the recent data  of STAR collaboration in
Au+Au collisions at RHIC with the statistical model results. Also
here the model is consistent with  data
\cite{xu,brauns}.\footnote{ A precise determination of $T_f$ is
not yet possible as fits with $170<T_f<190$   give similar value
of $\chi^2$. Data like eg. $\bar\Lambda /\bar p$ are necessary to
further determine $T_f$ \cite{brauns}. } The crucial difference to
the SPS is a much lower value of the baryon chemical potential
$\mu_B$ of 45 MeV. This is to be expected as higher collision
energy should lead to less stopping. The freezeout temperatures in
central A+A collisions at the SPS and RHIC coincide within errors
with the critical temperature from lattice QCD.  This could
indicate that all particles are originating from deconfined medium
and that the chemical composition of the system is established
during hadronization \cite{stachel,stoc,heinz1}.
\begin{figure}[h]
\vskip -1.6cm
\begin{minipage}[t]{80mm}
\includegraphics[ width=18.5pc, height=20.3pc]{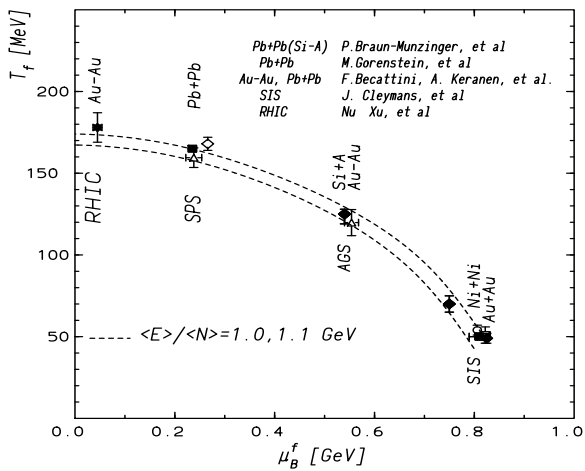}
\end{minipage}
\begin{minipage}[t]{74mm}
{\hskip 7.9cm \vskip -8.0cm {\caption {{Average temperature and
baryon chemical potential required to describe particle yields
measured at GSI/SIS, AGS, and SPS \cite{cleymans1,cleymans2}. The
unified chemical freezeout curve (dotted line) is calculated from
the condition that the ratio of the average energy $<E>$ per
average number of hadrons $<N>$ is equal to 1 or 1.1 GeV.  }}}}
\end{minipage}
\vskip -1.9 cm
\end{figure}

Chemical equilibration of secondaries is, however, not a unique
property of the SPS    data
\cite{hamieh,cleymans1,cleymans2,becattini2}. It is also present
at much lower incident energy or in peripheral high energy
nucleus-nucleus collisions  where the initial conditions required
for deconfinement are not necessarily established. To discuss
equilibration of strangeness in the above cases one needs,
however, to formulate a statistical model in the canonical (C)
ensemble with respect to the strangeness conservation
\cite{hamieh,cleymans2,hagedorn,b1}. \
\section{Chemical equilibration  - canonical description}
Strangeness conservation in statistical models can be described in
the (GC) ensemble only if the number of produced strange particles
per event is much larger than one. In the opposite limit of rare
particle production \cite{ko}, strangeness conservation must be
implemented locally on an event-by-event basis i.e.,   (C)
ensemble of strangeness conservation must be used. The (C)
ensemble is relevant in the statistical description of particle
production in low energy heavy ion \cite{cleymans1}, or high
energy hadron-hadron or $e^+e^-$ reactions \cite{b1} as well as in
peripheral heavy ion collisions \cite{hamieh}. The exact
conservation of quantum numbers, that is the canonical approach,
is known to reduce severely  the phase-space available for
particle productions. \cite{hagedorn}.
\begin{figure}[htb]
\vskip -0.5cm
\begin{minipage}[t]{80mm}
{\hskip -1.0cm
\includegraphics[width=24.2pc,height=15.5pc]{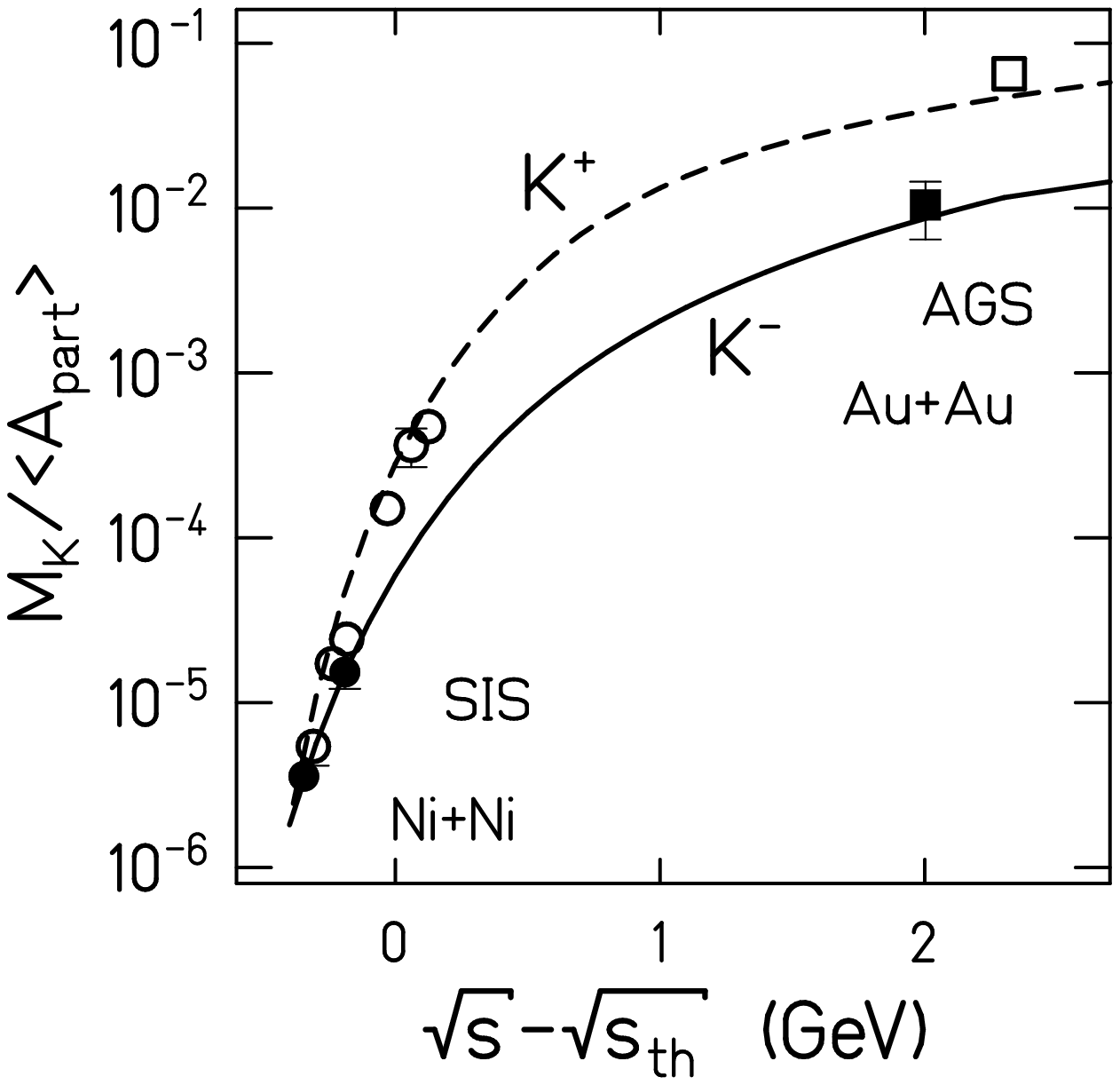}\\}
\vskip -1.2true cm \caption{Statistical model results on the
average number of kaons $M_K$ per participant $A_{part}$
\cite{cleymans2}. Open (full) symbols represent measured
multiplicities.}
\end{minipage}
\hspace{\fill}
\begin{minipage}[t]{75mm}
{\hskip -1.0cm
\includegraphics[width=24.2pc,
height=13.7pc]{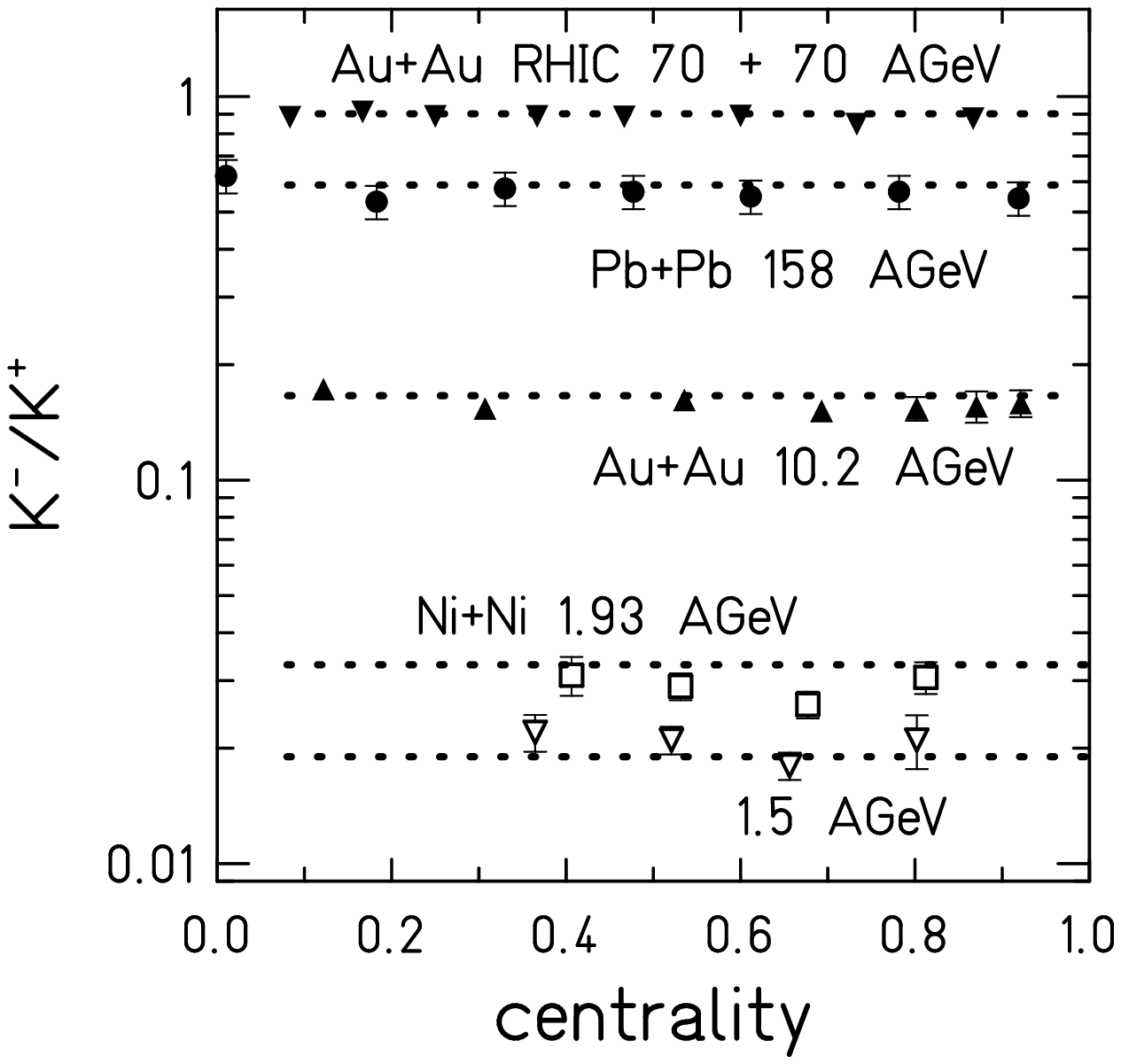}\\} \vskip -1.2 true cm \caption{$K^+/K^-$
data for different centrality at GSI/SIS  \cite{cleymans2}, AGS
\cite{ogilvie}, SPS \cite{sikler} and RHIC \cite{harris}. The
dotted lines are statistical model predictions \cite{cleymans2}. }
\end{minipage}
\end{figure}

To illustrate the canonical suppression let us consider a simple
example of $K^+K^-$ production in the pionic thermal system of the
volume $V$ at temperature $T$. For large  $T$, kaons are
abundantly produced and the density of kaons reaches (GC)
equilibrium value,
$n_{K^+}^{GC}=(1/2\pi^2)m_{K^+}^2TK_2(m_{K^+}/T)$. In the limit of
low temperature the $K^+,K^-$  are very rarely produced and in
order to satisfy strangeness neutrality condition, kaons must
appear in pairs in the near vicinity. In this case, the density of
produced kaons reaches an  equilibrium result of (C) ensemble. In
the asymptotic limit with  $<K><<1$ we have
\begin{equation}\label{1}
  n_{K^+}^C\sim [(1/2\pi^2)m_{K^+}^2TK_2(m_{K^+}/T)]\times
         [V_0(1/2\pi^2)m_{K^-}^2TK_2(m_{K^-}/T)].
\end{equation}
The first term coincides with  (GC) value,  the second describes
the phase space suppression since with each $K^+$ a $K^-$ has to
appear in the near vicinity  in order to conserve strangeness
exactly. The parameter $V_0$ in Eq. 1 is introduced  as a
correlation volume where the $K^+$ and $K^-$ should be created to
fulfill locality of strangeness conservation. In heavy ion
collisions $V_0$ was found to scale with $A_{part}$ and  
in proton induced precesses $V_0\sim V_{proton}$.
There are thus, two origins of canonical suppression of
strangeness: first, particles are produced in pairs what restricts
the available momentum phase-space  and secondly they appear  in
the near vicinity in space to fulfill locality of the conservation
laws.
\begin{figure}[htb]
\vskip -1.cm
\begin{minipage}[t]{80mm}
{
\includegraphics[width=17.5pc, height=15.3pc]{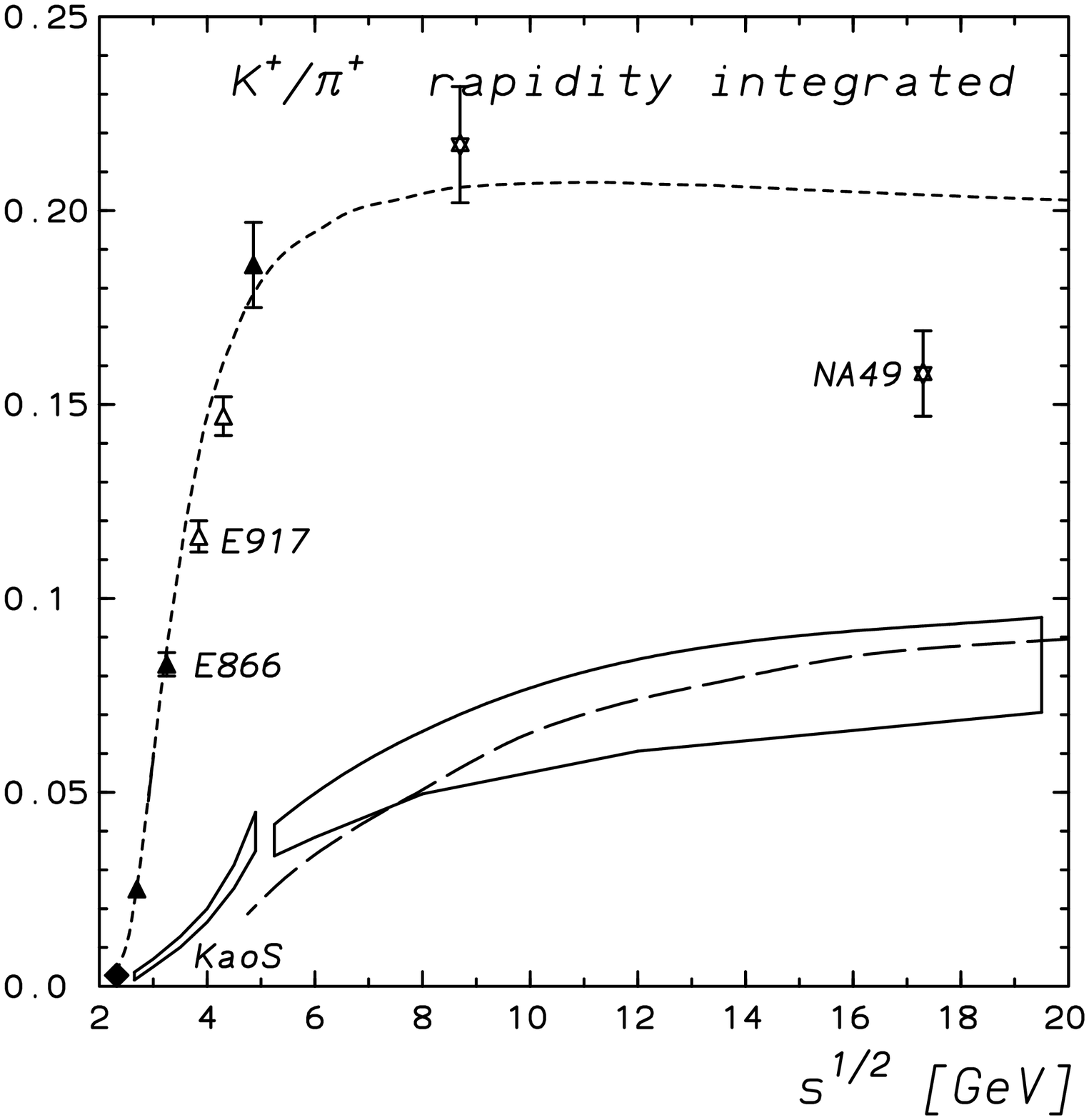}}\\
\vskip -2. true cm \caption{The $K^+/\pi^+$, 4$\pi$ ratio versus
energy \cite{blume}. The short-dashed and dashed lines  represent
thermal model result for Pb+Pb and p+p respectively. The full
lines are a parameterization of p+p data \cite{ogilvie}. }
\end{minipage}
\hspace{\fill}
\begin{minipage}[t]{75mm}
\includegraphics[width=17.5pc, height=15.2pc]{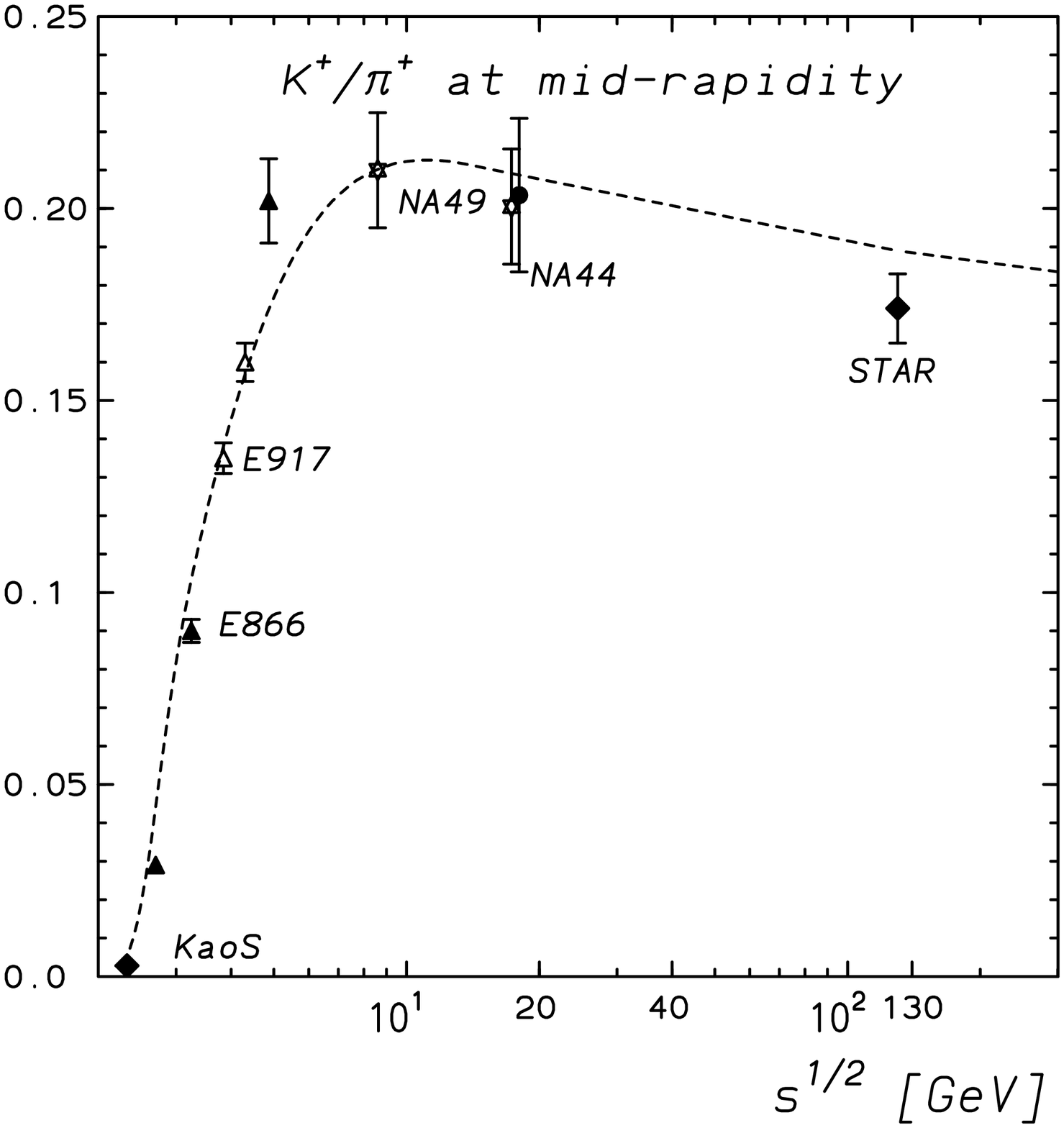}\\
\vskip -2. true cm \caption{As in Fig. 11 but  at midrapidity. The
results at RHIC were estimated from STAR results on $K^-/\pi^-$
and $K^+/K^-$ ratios \cite{harris}, assuming that the ratio of
$\pi^+/\pi^-\simeq 1$. }
\end{minipage}
\end{figure}
\subsection{Multistrange baryons - centrality dependence}

The canonical suppression of  thermal particle phase-space
increases with strangeness content of the particle. The exact
conservation of strangeness requires that each particle carrying
strangeness $\bar s$  has to appear e.g. with $s$ other particles
of the strangeness one to satisfy strangeness neutrality
condition.

In Fig. 6 we calculate the multiplicity/participant of $\Omega ,
\Xi ,$ and $\Lambda$   relative to its value in a small system
with only two participants. Thermal parameters  were assumed here
to be $A_{part}$ independent. Fig. 6 shows that the statistical
model in (C) ensemble  reproduces the basic features of  WA97
data: the enhancement pattern and enhancement saturation for large
$A_{part}$ indicating here  that  (GC) limit is reached. The
quantitative comparison of the model with experimental  data would
require an additional assumption on the variation of $\mu_B$ with
centrality to account for  larger value of $\bar B/B$ ratios in
p+A than in Pb+Pb collisions \cite{andersen}. An abrupt change of
the enhancement seen in the NA57 data for $\bar\Xi$ is, however,
{\it very unlikely} to be reproducible in terms of this approach.

One of the consequences  of the model is that the enhancement
pattern seen in Fig. 6 should   not be only   a  feature of the
SPS data. In terms of (C) model strangeness enhancement and
enhancement pattern should be also present there in heavy ion
collisions at lower energies and should be even more pronounced.
This is in contrast e.g. to UrQMD predictions which are showing
increasing
 enhancement with beam energy \cite{bleicher}.

In Fig. 7 we show a compilation of the data on $K^+/\pi^+$ ratio
in A+A relative to p+p collisions. This double ratio could be
referred to as a strangeness enhancement factor. The enhancement
is seen to be the largest  at the smallest  beam energy and is
smoothly decreasing towards higher energy. If  strangeness
enhancement is indeed of thermal origin then  similar behavior is
also expected for multistrange baryons. This could put in question
\cite{last} the observed strangeness enhancement  from p+p to A+A
collisions as an {\it appropriate characteristics} of the
deconfinement. To search for  the QGP formation through  the
strangeness composition of secondaries    one should rather look
for a  non-monotonic behavior of strange particle yields versus
centrality or collision energy.

A detailed analysis of the experimental data   in heavy ion
collisions  from SIS to SPS
\cite{braun,cleymans1,cleymans2,becattini2} has shown that most
particle yields are reproduced by the statistical
model.\footnote{Statistical models, however, failed e.g. to
describe the yield of $\eta$ at SIS or $\bar\Lambda /{\bar p}$
ratio at AGS.} In Fig. 8 we present the compilation of chemical
freezeout parameters required to reproduce measured particle
yields at SIS, AGS, SPS and RHIC energies. The GSI/SIS results
have the lowest freezeout temperature and the highest baryon
chemical potential. As the beam energy is increased a clear shift
towards higher $T$ and lower $\mu_B$ occurs. There is a common
feature of all these points namely that the average energy per
hadron is approximately 1 GeV. {\it Chemical freezeout} in A+A
collisions is thus reached {\it when the energy per particle drops
below 1 GeV} at all collision energies \cite{cleymans1}.
\begin{figure}[htb]
\vskip -0.9cm
\begin{minipage}[t]{80mm}
{\hskip -.1cm
\includegraphics[width=18.1pc, height=15.1pc]{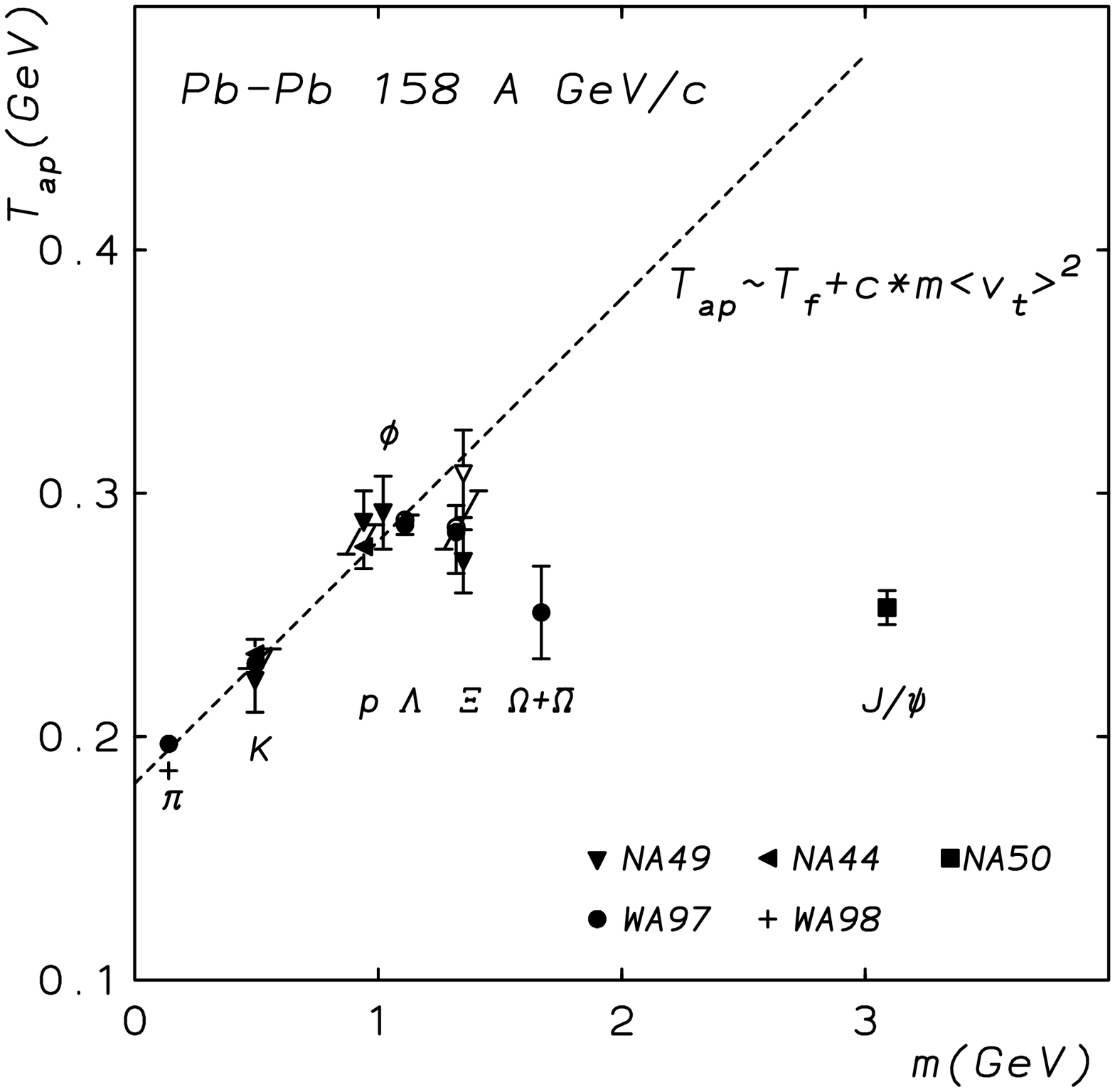}}\\
\vskip -2.3true cm \caption{ Dependence of the transverse mass
inverse slope parameters on the particle rest mass
\cite{hecke,na50}.}
\end{minipage}
\hspace{\fill}
\begin{minipage}[t]{75mm}
{\hskip -.1cm
\includegraphics[width=18.1pc, height=15.1pc]{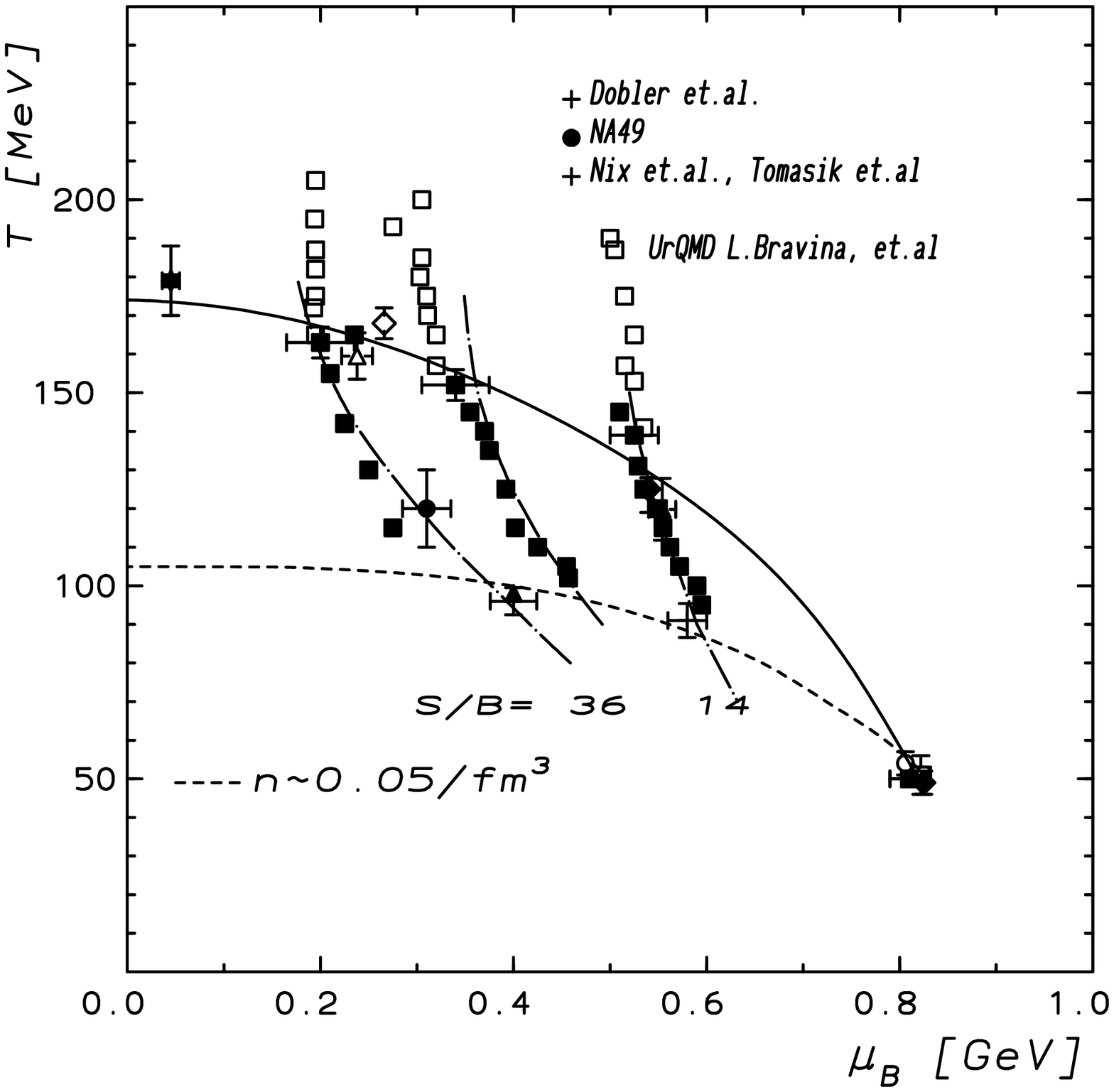}}\\
\vskip -2.3 true cm \caption{Thermal \cite{heinz1,tomasik}  and
chemical freezeout conditions \cite{cleymans1,cleymans2} compared
with the evolution path from UrQMD \cite{bravina}.        }
\end{minipage}
\vskip -0.7true cm
\end{figure}
Fig$^,$s. 9-12  are showing  different particle multiplicity
ratios along unified freezeout curve  in comparison with
experimental data. In all cases the energy dependence of data is
well reproduced by the statistical model.  The canonical
strangeness suppression in central A+A collisions is very relevant
at SIS energies.  At the top AGS the (GC) treatment of strangeness
conservation is already adequate.

Specific features of $K^+$ and $K^-$  excitation functions shown
in Fig. 9 are appearing in the model as a consequence of different
dependence of $K^+$ and $K^-$ yields on baryon number density
\cite{cleymans2}. Also here, the increase of the canonical
correlation volume from the size of the colliding nucleus at SIS
up to the macroscopic volume of a thermal fireball at  SPS is of
crucial importance. Furthermore, the independence of the $K^+/K^-$
ratio on the number of participating nucleons seen in Fig. 10 from
SIS to RHIC is a natural result of the statistical model as a
particle and its antiparticle have the same centrality dependence
as well as in (GC) and  in (C) formulation. A consistent
description of the energy or centrality dependence of kaon yields
in A+A collisions within microscopic transport models is still
missing \cite{cassing}.

Of particular interest are experimental results represented in
Fig.11 where the 4$\pi$ data on $K^+/\pi^+$ ratios \cite{blume}
are showing a maximum at $\sqrt s\sim 6-8$GeV and drop at the SPS.
The suppression of this ratio disappears, however, at midrapidity
\cite{blume} as seen in Fig. 12. Statistical model favors the
behavior seen in Fig. 12. The description of fully integrated data
at the top SPS energy   is only possible in terms of the
statistical model when introducing an extra  parameter which
accounts for strangeness undersaturation \cite{becattini2}. The
suppression of $K^+/\pi^+$ ratio seen in  Fig. 11 was predicted as
a possible signal for deconfinement \cite{gazdzicki}.

\section{From chemical to thermal freezeout}

Transverse mass distribution of strange and non strange particles
clearly indicates in addition to the thermal one also  the
collective transverse flow component \cite{stoc,heinz1,tomasik}.
In Fig.13 the resulting increase of the transverse mass slope
parameter $T_{app}$ with the rest mass of the particle is  typical
for transversely expanding source. Deviations of $\Omega$
\cite{hecke} and $J/\Psi$ seen in Fig.13 most likely reflect their
earlier decoupling from the system due to a small rescattering
cross section with the surrounding medium.

 The compilation of thermal
freezeout parameters from SIS to SPS,  extracted from particle
spectra and two particle momentum correlation   are shown in Fig.
13.  The system seems to expand transversely  from the chemical to
thermal freezeout following  the path of constant entropy (S) per
baryon (B) \cite{cleymans1}. The abundances of secondaries during
this evolution are frozen in. In  the statistical approach this
can only happen if the phase space of  particles like e.g. pion ,
kaon and nucleon are over-saturated \cite{rapp} by an effective
chemical potentials with typical values of 60-120MeV at
$T_{th}\sim 110$MeV . The evolution path and the moment  of
dynamical decoupling of the fireball  is seen in Fig. 14 to
coincide with UrQMD description \cite{soff,bravina}. The chemical
freezeout curve of constant $<E>/<N>\simeq 1$GeV appears in UrQMD
\cite{bravina} when the system reaches the kinetic equilibrium
(full squares in Fig. 14).

\section{Conclusions}
One of the most puzzling  results in heavy ion collisions is
presumably  an observation  that particles seem to be produced
according to the principle of maximal entropy. In a very broad
energy range  from $\sqrt s\sim 2.5 $  up to 130 GeV hadronic
yields and their ratios observed in heavy ion collisions resemble
chemical equilibrium population along   unified freezeout curve
determined by the conditions  of fixed energy/hadron $\simeq
1$GeV. Strangeness follows this systematics and since their
production mechanism is quite different at low and  at high
energies it seems that particle yields are losing information
about dynamics. However, there are some characteristic features of
the system at chemical freezeout in central A+A collisions being
only present at SPS and RHIC: i) chemical freezeout temperature is
within errors consistent with the critical temperature  and ii)
strangeness is un-correlated and redistributed in the whole volume
of a thermal fireball. Although the condition  i) is also there in
peripheral nucleus-nucleus and even in p+p collisions, the
property ii) is obviously  not valid any more. Here, the thermal
phase space available for strange particles is strongly suppressed
since, for only a few particles being produced per event,
strangeness is strongly correlated in the volume $V_0\sim
V_{proton}$ and thus has to be conserved exactly and locally
(canonical description). In high energy A+A reactions, there are
already sufficiently many strange particles being produced per
event such that  strangeness is conserved on the average and is
distributed in the whole volume of the fireball (grand canonical
description).

If  thermal description of particle productions  is correct,
strangeness enhancement from p+p to A+A collisions appears
naturally as a transition from  the canonical  to the grand
canonical regime. Hence strangeness  enhancement does not
necessarily require deconfinement and should also be observable at
lower collision energies where the conditions for deconfinement
are not satisfied. However, a specific, non-monotonic behavior of
strangeness versus centrality or collision energy could be of
interest  in this context. Recent results of NA57 and NA52 showing
an abrupt change with centrality of $\bar\Xi$ and $K^+$,
respectively, as well as the observed by NA49 drop of $K^+/\pi^+$
ratio  seen in Fig. 11 could signal a new dynamics.

\subsection*{Acknowledgments}
\hspace*{\parindent} We  acknowledge  stimulating discussions with
H. Bia\l kowska, Ch. Blume, B. Friman, P. Braun-Munzinger, H.
Satz, E. Shuryak, H. Specht, R. Stock and N. Xu.  Discussions with
H. Oeschler, A. Tounsi and the co-authors of \cite{becattini2} are
also acknowledged.


\begin{thebibliography}{9}

\bibitem{qm}
          {Proceedings of  {\it Quark Matter '99 }},
  Nucl. Phys. A661 (1999).

\bibitem{satz} H. Satz,  Rept. Prog. Phys. 63 (2000) 1511.

\bibitem{stachel} P. Braun-Munzinger and J. Stachel, Nucl. Phys. A606 (1996) 320;
J. Stachel, Nucl. Phys.  A654 (1999)  119c.

\bibitem{stoc} R. Stock,  Phys. Lett. 456 (1999) 277;
 Prog. Part. Nucl. Phys.  42 (1999) 295.

\bibitem{heinz1} U. Heinz, Nucl. Phys.  A685 (2001) 414;
 Nucl. Phys. A661 (1999) 349.

\bibitem{karsch} F. Karsch, Nucl.Phys. B (Proc. Suppl.) 83-84
(2000) 14 and {\it these proceedings}.

\bibitem{rafelski1} J. Rafelski and B. M\"uller, Phys. Rev. Lett. 48
(1982) 1066; P. Koch, B. M\"uller and J. Rafelski, Phys. Rep. 142
(1986) 167; J. Rafelski Phys. Lett. B262 (1991) 333.


\bibitem{letessier} J. Letessier and J. Rafelski,
Int. J. of Mod. Phys. E9 (2000) 107.


\bibitem{sikler} F. Sikler, $et$ $al.$, NA49 Collaboration,
Nucl. Phys.  A661, (1999) 45c {\it and ref. therein}.

\bibitem{bialkowska} H. Bialkowska, $et$ $al.$, NA35
Collaboration, Z. Phys. C64 (1994) 381.


\bibitem{andersen} E. Andersen, $et$ $al.$, WA97 Collaboration,
Phys. Lett. B449 (1999) 401.

\bibitem{carrer} N. Carrer, NA57 Collaboration, {\it these
proceedings}.


\bibitem{kabana} S. Kabana, {\it et al.}, NA52 Collaboration, J.
Phys. G27 (2001) 495.

\bibitem{soff} S. Soff, $et$ $al.$, J. Phys.G27 (2001) 449.

\bibitem{bravina}  L. Bravina, {\it these proceedings}.

\bibitem{vance} S. E. Vance, {\it et al.}, Phys. Rev. Lett. 83
(1999) 1735; J. Phys. G27 (2001) 627.

\bibitem{bleicher} M. Bleicher, W. Greiner, H. St\"ocker and N. Xu,
Phys. Rev.C62 (2000) 061901.

\bibitem{capella} A. Capella and C. A. Salgado, Phys. Rev. C60
(1999) 054906.

\bibitem{lin} Z. Lin, {\it et al.}, nucl-th/0011059 and {\it these
proceedings}.

\bibitem{braun} P. Braun-Munzinger, I. Heppe and J. Stachel, Phys.
Lett.  B465, (1999) 15.

\bibitem{hamieh}J. S. Hamieh, K. Redlich and A. Tounsi,
Phys. Lett. { B486} (2000) 61.

\bibitem{xu} N. Xu, STAR Collaboration, {\it these proceedings}.


\bibitem{harris} J. Harris, STAR Collaboration, {\it these proceedings}.


\bibitem{brauns} P. Braun-Munzinger, D. Magestro and J. Stachel, {\it  to appear}.

\bibitem{cleymans1} J.  Cleymans, {\it et al.}, Phys. Rev.C60 (1999)
054908; Phys. Rev. Lett. 81, (1998) 5284.

\bibitem{cleymans2} J.  Cleymans,  {\it et al.},
Phys. Rev. C59 (1999) 1663; Phys. Lett. B485 (2000) 27.


\bibitem{becattini2} F. Becattini, J. Cleymans, A. Keranen, E.
Suhonen and K. Redlich, nucl-ph/0011322.



\bibitem{hagedorn}R. Hagedorn, CERN Rep. 71 (1971); E.
Shuryak, Phys. Lett. B42 (1972) 357; J. Rafelski,  Phys. Lett. B97
(1980) 279; R. Hagedorn, {\it et al.}, Z. Phys. { C27} (1985) 541.

\bibitem{b1}
F. Becattini,  Z. Phys. C69 (1996) 485; F. Becattini, {\it et
al.}, Z. Phys. C76 (1997)  269.

\bibitem{ko}  C.M. Ko, V. Koch, Z. Lin, K. Redlich,
M. Stepanov and X.N. Wang, nuc-th/0010004.

\bibitem{last} {\it See also:} H. Drescher, J. Aichelin and K. Werner, Rapport, Subatech, 00-21.

\bibitem{hecke} H. van Hecke, H. Sorge and N. Xu, Nucl. Phys. A661
(1999) 493c.

\bibitem{ogilvie} J. C. Dunlop {\it et al.}, Phys. Rev. C61
(2000) 031901; C. A. Ogilvie, {\it these proceedings}.

\bibitem{blume} Ch. Blume, NA49 Collaboration, {\it these
proceedings}.


\bibitem{cassing} W. Cassing, Nucl. Phys. A661 (1999) 468c and {\it references
therein}.


\bibitem{gazdzicki} M. Gazdzicki, {\it et al.}, Z. Phys. C65 (1995) 215,
Acta Phys. Pol.  B30 (1999) 2705.

\bibitem{tomasik} B. Tomasik, {et al.}, Nucl. Phys. A663 (2000)
753c.

\bibitem{rapp} R. Rapp and E. Shuryak, hep-ph/0008326, E. Suryak,
{\it et al., to appear}.

\bibitem{greiner} C. Greiner and S. Leupold, nucl-th/0009036, C.
Greiner, {\it these proceedings}.

\bibitem{na50} By NA50 Collaboration, Phys. Lett. B499 (2001) 85.


\end{thebibliography}
\end{document}